\documentclass[conference]{IEEEtran}
\IEEEoverridecommandlockouts
\usepackage{cite}
\usepackage{amsmath,amssymb,amsfonts}
\usepackage{footmisc}
\usepackage{algorithmic}
\usepackage{graphicx}
\usepackage{textcomp}
\usepackage{color,soul}
\usepackage{xcolor}
\usepackage{amsmath}
\usepackage{fancyhdr}

\makeatletter
\newcommand*\titleheader[1]{\gdef\@titleheader{#1}}
\AtBeginDocument{%
  \let\st@red@title\@title
  \def\@title{%
    \bgroup\normalfont\large\centering\@titleheader\par\egroup
    \vskip1.5em\st@red@title}
}
\makeatother

\def\BibTeX{{\rm B\kern-.05em{\sc i\kern-.025em b}\kern-.08em
    T\kern-.1667em\lower.7ex\hbox{E}\kern-.125emX}}

\usepackage{xcolor, pifont}
\newcommand*\colourtick[1]{%
  \expandafter\newcommand\csname #1tick\endcsname{\textcolor{#1}{\ding{51}}}%
}
\newcommand*\colourcross[1]{%
  \expandafter\newcommand\csname #1cross\endcsname{\textcolor{#1}{\ding{55}}}%
}
\colourtick{blue}
\colourcross{red}

\makeatletter
\newcommand{\ssymbol}[1]{^{\@fnsymbol{#1}}}
\makeatother

\title{Robust Image Watermarking based on Cross-Attention and Invariant Domain Learning}

\titleheader{}

\author{
    \IEEEauthorblockN{
    Agnibh Dasgupta\IEEEauthorrefmark{1}, 
    Xin Zhong\IEEEauthorrefmark{1}}\
    
    \smallskip
    \IEEEauthorblockA{
    \IEEEauthorrefmark{1}
    Department of Computer Science, University of Nebraska Omaha, Omaha, NE, USA
    \\ 
    adasgupta@unomaha.edu, xzhong@unomaha.edu
    }
}

\begin{document}

\maketitle

\begin{abstract}
Image watermarking involves embedding and extracting watermarks within a cover image, with deep learning approaches emerging to bolster generalization and robustness. Predominantly, current methods employ convolution and concatenation for watermark embedding, while also integrating conceivable augmentation in the training process. This paper explores a robust image watermarking methodology by harnessing cross-attention and invariant domain learning, marking two novel, significant advancements. First, we design a watermark embedding technique utilizing a multi-head cross attention mechanism, enabling information exchange between the cover image and watermark to identify semantically suitable embedding locations. Second, we advocate for learning an invariant domain representation that encapsulates both semantic and noise-invariant information concerning the watermark, shedding light on promising avenues for enhancing image watermarking techniques.

\end{abstract}


\section{Introduction}
\label{sec:Introduction}
\thispagestyle{fancy}
\renewcommand{\headrulewidth}{0pt}
\fancyfoot[R]{The source code, pretrained model, and additional results are available at~\\\texttt{https://github.com/cent664/SSRIW}.}

Digital image watermarking is a method in which watermark information is subtly embedded within digital images. This technique is employed for a myriad of purposes aimed at safeguarding digital intellectual property, including content or owner authentication, copy control, and copyright management. 
Traditional image watermarking is often hampered by the necessity of manual design, as each instance necessitates a bespoke design to accommodate varying watermarks and distortions~\cite{Trad1, Trad2}. 
This manual process can be labor-intensive and may not always yield optimal results. On the other hand, the advent of deep neural networks has paved the way for learning generalizable complex mappings, which significantly mitigates these challenges. 
Consequently, deep learning-based watermarking has surfaced as a promising alternative to the traditional watermarking methodologies, particularly for images~\cite{DL1, DL2, DL3, DL4}, but also extending its utility to audio\cite{Arjon} and video domains~\cite{video_attention}~\cite{video_framework}. 
The deep learning models can autonomously learn and adapt to various watermarking requirements and distortions, potentially offering a more robust and efficient solution for protecting digital assets across different media.

A prevalent method for deep learning-based watermark embedding involves training a Convolutional Neural Network (CNN) to extract pertinent features from both the cover image and the watermark, followed by concatenating these features together. 
The combined features are then utilized to generate the watermarked image. 
Nonetheless, recent advancements have shown that Vision Transformers (ViTs)~\cite{ViT} are more adept for tasks necessitating a comprehensive understanding of an image's content as a whole. 
This is attributed to their capability to grasp global contextual information through an attention mechanism, a stark contrast to CNNs, which excel in local feature extraction~\cite{dosovitskiy2021image}. 
The global perspective afforded by ViTs can potentially lead to more coherent and effective image watermarking, especially in scenarios where understanding the overall content of the image is crucial for embedding watermarks intelligently and inconspicuously. 
However, utilizing the attention ability of ViTs for image watermarking remains unexplored.

   
   

Furthermore, the integrity and robustness of the extracted watermark can be substantially compromised by intentional or unintentional distortions inflicted on the watermarked image, owing to the fragile nature of deep learning models in handling such distortions~\cite{robustness_watermarking}. 
Although various approaches have been proposed to address this robustness issue~\cite{HiDDeN, Zhong, Distortion_agnostic, Cover_agnostic}, a predominant portion of current solutions hinge on incorporating conceivable distortions into the training pipeline~\cite{zhong2023brief}. 
Consequently, the attained robustness predominantly pertains to the specific distortions that have been integrated during training~\cite{zhong2023brief}, leaving the system potentially vulnerable to unforeseen or novel distortions. 
This methodological limitation underscores a potential avenue for advancing the robustness of deep learning-based watermarking systems against a broader spectrum of distortions, thereby enhancing their reliability and applicability in real-world scenarios. 
To achieve this objective, a promising avenue is the training of effective invariant domain through self-supervised learning (SSL). 
While SSL is often lauded for its representative features, the acquisition of nearly identical representations of two augmentations (\textit{i.e.}, the invariance) is integral to its common training paradigms~\cite{Hinton, LeCun}. 
Certain existing methodologies have ventured into leveraging this invariant domain for image watermarking tasks~\cite{facebook_DINO}. 
Nonetheless, prevailing methods tend to regard SSL models as ready-made, and the concurrent training of invariant domain alongside image watermarking remains an uncharted territory.

In this paper, we introduce a robust image watermarking method leveraging cross-attention and invariant domain learning, unfolding two significant novel avenues. 
Firstly, we devise an image watermarking embedding technique through the implementation of a multi-head cross attention mechanism, facilitating an information interchange between the cover image and watermark to discern semantically apt embedding locales. 
Secondly, we propose to learn an invariant domain encapsulating both semantic and noise-invariant information regarding the watermark.
To achieve this invariant domain, we develop a self-supervised watermarking framework that simultaneously learns watermarking and invariant domain from scratch.  
This exploration in training methodology lays the groundwork for enhancing robustness in deep learning-based image watermarking. 
Experimentally, we discern that our advocated method either matches or surpasses robustness across various noise scenarios compared to state-of-the-art techniques.

The remainder of the paper is structured as follows: Related work is explored in Section~\ref{sec:Related Work}, while the intricacies of the proposed watermarking scheme are elucidated in Section~\ref{sec:The Proposed Scheme}. Data, experimental setups, and analyses are delineated in Section~\ref{sec:Experiments}, and finally, conclusions derived from our investigations are deliberated in Section~\ref{sec:Conclusion}. 

\section{Related Work}
\label{sec:Related Work}
\subsection{Image Watermarking and Deep Learning}
Traditional image watermarking can be classified into spatial and frequency domain based approaches, both of which are computationally inexpensive but come with the constraint of manual design~\cite{Trad1, Trad2}. 
Since Zhu \textit{et al}. introduced the concept of joint training in their end-to-end neural network based watermarking scheme HiDDeN~\cite{HiDDeN}, the generalization power of deep neural networks has been explored extensively in image watermarking. 
Much of the following research has been focused on expanding on their embedder-extractor joint training by covering up limitations such as the need for a differentiable noise layer and lack of robustness to noises not included in the noise layer. 
To deal with the need for gradient flow, Liu \textit{et al}.~\cite{two_stage} proposed a two-stage training scheme where they disentangle the embedder and extractor training, allowing smooth gradient flow in the embedding stage and move the focus on training the extractor with noise. 
Zhong \textit{et al}.~\cite{Zhong} introduced an invariance layer design to remove redundant information when extracting the watermark and use multi-scale inception networks to get an intricate embedding of the watermark while maintaining imperceptibility. 
Xu \textit{et al}.~\cite{INN} employed invertible neural networks using their bijective propagation to improve watermarking embedding and extracting to improve robustness and embedding quality. 
Deep learning based watermarking has also been applied audio~\cite{Arjon} and video~\cite{video_attention, video_framework}, demonstrating their superior robustness over traditional watermarking schemes. 
While a wealth of research focuses on enhancing the quality and robustness of watermarking, hardly any delves into the cross attention mechanism of ViTs, which could potentially benefit image watermarking through its global analysis capability.

\subsection{Image Watermarking and the Invariant Domain}
In pursuit of acquiring representative features, numerous contrastive learning approaches, \textit{e.g.}, the studies underscored by Chen \textit{et al}.\cite{Hinton} and Bardes \textit{et al}.\cite{LeCun}, have employed the notion of learning generalizable invariant image domains. 
In their methodologies, the output features for an image and its transformed variant are derived from joint-embedding architectures. These `positive' outputs derived from the same image are then drawn closer to each other, while other negative samples and their transformations are distanced to achieve an invariant latent domain. 

Vukoti\'c \textit{et al}.~\cite{vukotic2020classification} initially embed watermarks into the representations gleaned from readily available convolutional models. Subsequently, recognizing the promise of SSL for robust domain, Fernandez \textit{et al}.~\cite{facebook_DINO} employed the pre-trained SSL model DINO~\cite{DINO} to generate a domain presumed to be invariant. 
They embed watermarks into these features with the methods described in~\cite{vukotic2020classification}, demonstrating extraction performance on par with contemporary state-of-the-art approaches.
Nevertheless, as current methodologies treat SSL models as ready-to-use, the concerted exploration of training image watermarking alongside an invariant domain remains unexplored.


\section{The Proposed Scheme}
\label{sec:The Proposed Scheme}
\subsection{Architecture}
\label{sec:Architecture}

\begin{figure*}[!htb]
    \centering
    \vspace{-1.5em}
    \includegraphics[width=0.95\linewidth]{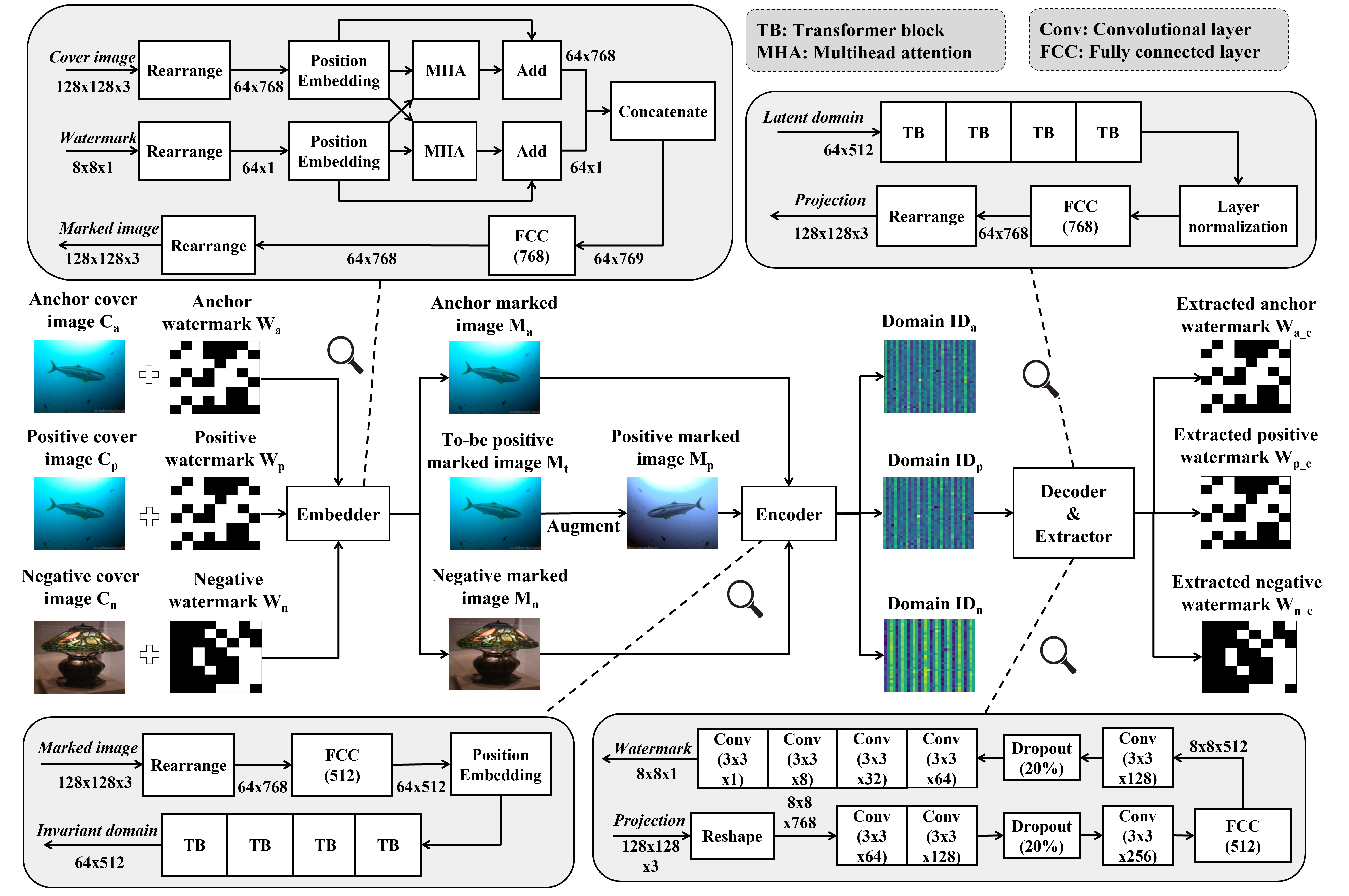}
    \vspace{-1.0em}
    \caption{The proposed method.}
    \label{fig:Architecture}
    \vspace{-1.5em}
\end{figure*}

The proposed architecture comprises four main components: the embedder, encoder, decoder, and extractor. The components are trained in the same order, the details of which are explained in the following section~\ref{sec:Training scheme}.

First, we outline the terminology and workflow depicted in Figure~\ref{fig:Architecture}. The symbols $C_{a}$, $C_{p}$, and $C_{n}$ denote the anchor, positive, and negative cover images, while $W_{a}$, $W_{p}$, and $W_{n}$ represent their corresponding watermarks. Initially, an anchor cover image $C_a$ and watermark $W_a$ are input into our embedder, yielding a marked image $M_a$. Similarly, $M_p$ and $M_n$ are obtained from pairs $(C_p, W_p)$ and $(C_n, W_n)$, respectively. These marked images are fed into our encoder, producing invariant domains $ID_{a}$, $ID_{p}$, and $ID_{n}$, abstracting essential features while mitigating noise effects. The invariant domains are then input into our decoder and extractor, resulting in the extracted watermarks $W_{a\_e}$, $W_{p\_e}$, and $W_{n\_e}$. These extracted watermarks should closely resemble the original watermarks, showcasing the robustness of our scheme. The flow from cover images and watermarks, through the embedding, encoding, and extraction processes, to the extracted watermarks, encapsulates our proposed watermarking scheme, as shown in Figure~\ref{fig:Architecture}.

Other than the extractor, which primarily consists of convolutional layers (\textit{kernel size * kernel size * number of output filters}) and fully connected layers, the other three components adopt the structures from ViT.
The transformer blocks (TB) and multihead attention (MHA) layers utilize a channel dimension, featuring a hidden layer of size 512 with two heads. The MHA employs patch sizes of 16 and 1 for the cover image and watermark, respectively. 

\subsubsection{Embedder}
\label{sec:Embedder}
The embedder net fuses a $128 \times 128 \times 3$ cover image $C$ with an $8 \times 8 \times 1$ watermark $W$ into a $128 \times 128 \times 3$ marked image $M$, such that $C$ and $M$ are visually identical. The embedder therefore takes two inputs of shape $128 \times 128 \times 3$ and $8 \times 8 \times 1$ and learns a mapping to one output of shape $128 \times 128 \times 3$. Both $C$ and $W$ are rearranged to vectors consisting of $16 \times 16$ and $1 \times 1$ patches respectively, and these sequences of patches are then embedded with positional information~\cite{ViT} and passed through separate multihead cross-attention layers with a fully-connected layer of 512 units and 2 heads. These attention outputs obtained are added to the original patch embeddings and are then concatenated along the channel dimension to obtain a $128 \times 128 \times 784$ feature map. After passing through a channel-wise fully connected layer with 768 units, the patch map is rearranged back to an output shape of $128 \times 128 \times 3$.

\subsubsection{Encoder and Decoder}
\label{sec:Encoder and Decoder}
The encoder takes a $128 \times 128 \times 3$ marked image $M$ as input and outputs a compressed latent domain $ID$ containing semantic information about $M$. The input is first rearranged to a vector of $16 \times 16$ patches, and then this vector is passed through a channel-wise fully connected layer with 512 units, followed by embedding with positional information. The patch embeddings are then passed through four transformer blocks~\cite{ViT}, and an output vector of patches is obtained.

The decoder takes the latent domain $ID$ as input and outputs a vector of shape $128 \times 128 \times 3$. Its job is to assist the extractor with recovering the watermark from the invariant domain by projecting the vector of patches to the same shape as the marked image. As with traditional autoencoder architectures, the decoder is essentially the encoder reversed. Therefore, the input sequence of patch vectors is first passed through four transformer blocks. The output is then passed through a channel-wise fully connected layer with 512 units and rearranged back to the shape $128 \times 128 \times 3$.

\subsubsection{Extractor}
\label{sec:Extractor}
The extractor takes a $128 \times 128 \times 3$ vector and outputs an $8 \times 8 \times 1$ watermark $W'$, such that $W$ and $W'$ are as similar as possible. $M$ is first reshaped to $8 \times 8 \times 48$ to fit the watermark shape. Then it is passed through a series of three convolutional layers for feature extraction, with 64, 128, and 256 filters respectively. After this feature space expansion, it is passed through a channel-wise fully connected layer with 512 units and finally reduced channel-wise through a sequence of convolutional layers with 128, 64, 32, 8, and 1 filters to get $W'$. To ensure gradient stability during training, dropout layers (20\%) are incorporated into the architecture.

\subsection{Benefits of the Cross Attention (MHA)}
\label{sec:Cross attention in watermarking}
The emergence of ViTs has altered the paradigm in feature extraction and representation learning, outperforming state-of-the-art CNNs on multiple benchmarks with fewer training computational resources~\cite{MDPI_ViTvsCNN}. ViTs utilize a self-attention mechanism to efficiently capture global context among pixels, aiding in understanding spatial relationships within images. Unlike self-attention which focuses on elements within a sequence, cross-attention facilitates information exchange between different sequences. The application of ViT in image watermarking remains largely unexplored.

\begin{figure}[!htb]
    \centering
    \vspace{-1.0em}
    \includegraphics[width=0.99\linewidth]{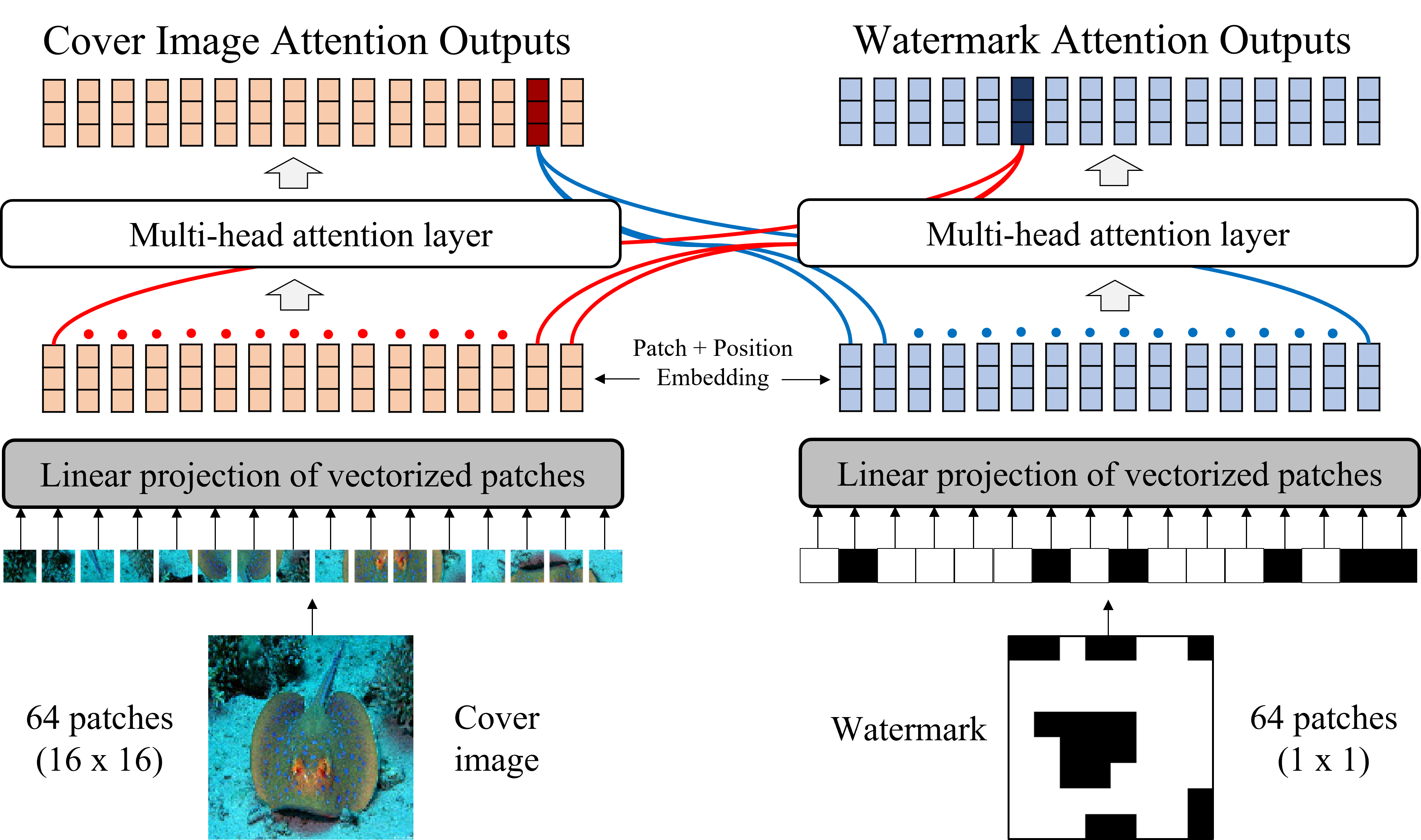}
    \caption{Proposed cross attention for watermark embedding.}
    \label{fig:cross_attention_watermarking}
    \vspace{-0.5em}
\end{figure}

We propose utilizing MHA in image watermarking to allocate watermarks across various regions based on relevance. The method employs self-attention and cross-attention for intra-image and inter-image relationships, respectively. Self-attention divides an image into patches, computing attention scores among them to capture contextual relationships. By treating the watermark as auxiliary information and applying cross-attention between image and watermark embeddings, we aim for a more robust and imperceptible watermarking process.
Figure~\ref{fig:cross_attention_watermarking} illustrates the decomposition of the cover image and watermark into vectors, each comprising 64 patches, with shapes of $16 \times 16$ and $1 \times 1$ respectively. Post positional embeddings augmentation, these vectors are processed through an MHA layer, computing attention scores between patches from one image (as queries) and the other (as keys), facilitating understanding between the cover image and the watermark. These scores are then utilized to identify optimal watermark embedding locations.

\subsection{The Invariant Domain}
\label{sec:Invariant Domain}
The invariant domain aims to encapsulate essential patterns and salient details from the marked image, serving as a robust anchor for watermark embedding, and thereby ensuring resiliency against various distortions. By successfully capturing these features, robust watermark extraction can be achieved even amidst heavy distortions. Our proposed method learns this invariant domain by minimizing the distance between the encodings of $M_a$ and $M_p$, while maximizing the distance between $M_a$ and $M_n$.


The triplet approach utilized in our model is elaborated in Section~\ref{sec:Loss}. This method facilitates the learning of common features shared between the anchor and the positive image, aiming to capture the common watermark since $M_a$ and $M_p$ have the same watermark, in a manner consistent across similar images. The inclusion of a negative image acts as a regularizer, preventing model collapse by encouraging distinction between relevant and irrelevant features for watermark embedding. This way, the model is guided to recognize and utilize the intrinsic relationships among the images, ensuring robust watermark embedding and extraction, even amidst potential distortions or variations in the image data. The triplet learning framework significantly enhances the robustness and learning capability of our watermarking scheme, adeptly handling watermarking scenarios with diverse image data.

\subsection{Training scheme of the Proposed Framework}
\label{sec:Training scheme}

\begin{figure}[!ht]
    \centering
    \vspace{-0.5em}
    \includegraphics[width=1.0\linewidth]{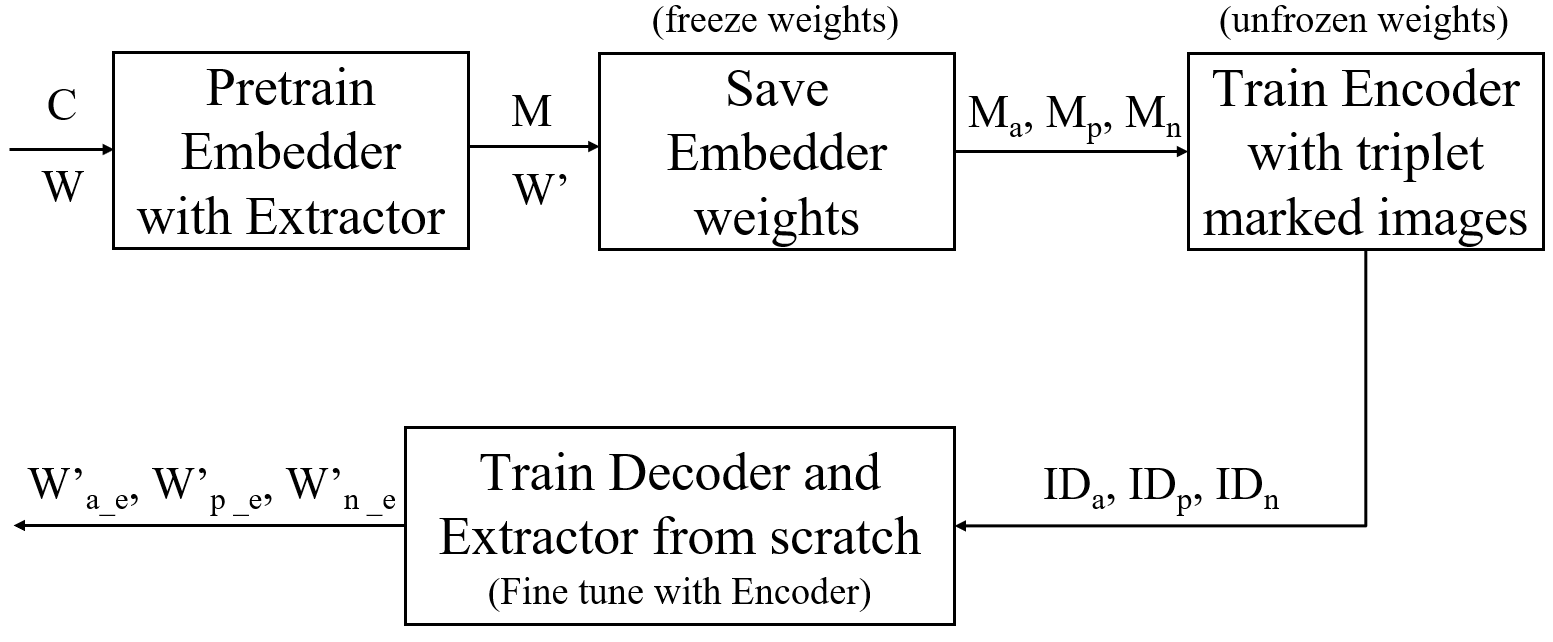}
    \vspace{-1.5em}
    \caption{The training flow of our proposed scheme.
    }
    \vspace{-0.5em}
    \label{fig:training_flow}
\end{figure}

We propose a self-supervised watermarking training scheme that simultaneously learns watermarking and invariant domain
from scratch, as detailed in Figure~\ref{fig:training_flow}.

\subsubsection{Embedder-Extractor Pretraining} 
The embedder and extractor are pretrained together as a pair in isolation without any noise. 
The embedder takes two inputs, cover image $C$ and watermark $W$, and has a single output, marked image $M$. 
Its goal is to generate a marked image as similar as possible to the cover image while containing the given watermark information. 
The extractor takes one input, marked image $M$, and outputs a single watermark $W'$. 
Its job is to extract the watermark $W'$ from the marked image such that it is as similar as possible to the originally embedded watermark $W$.

While the embedder-extractor is trained using a single cover image and watermark pair, during the rest of the training scheme, three pairs are used. They are denoted as ($C_{a}$, $W_{a}$), ($C_{p}$, $W_{p}$), and ($C_{n}$, $W_{n}$) and output marked images $M_{a}$, $M_{p}$, and $M_{n}$ respectively.

\subsubsection{Encoder training}
After pretraining, the extractor weights are discarded and the embedder weights are frozen.
$M_{t}$ is augmented with compound noises as described in section~\ref{sec: Data augmentation} to obtain $M_{p}$. 
The encoder is then trained in isolation using triplet loss, which takes a triplet input: anchor image $M_{a}$, positive image $M_{p}$, negative image $M_{n}$, and attempts to output an invariant domain for each image, $ID_{a}$, $ID_{p}$, and $ID_{n}$. 
These inputs are obtained from the output of the pretrained embedder network. 
The encoder's goal is to make $ID_{a}$ and $ID_{p}$ as similar as possible, and $ID_{a}$ and $ID_{n}$ as different as possible while preserving semantic information about the images using triplet loss.

\subsubsection{Decoder and Extractor training}
Once the encoder is trained, its weights are unfrozen and it is fine-tuned along with the decoder and extractor. 
The output from the encoder is passed through the decoder, reshaping it to $128\times128\times3$, and is then passed through the extractor. 
Both the decoder and extractor are trained from scratch to obtain the watermarks from $ID_{a}$, $ID_{p}$, and $ID_{n}$. 
The outputs, therefore, are $W_{a\_e}$, $W_{p\_e}$, and $W_{n\_e}$.

\subsection{Loss Computation}
\label{sec:Loss}


Triplet loss utilizes a triplet comprising an 'anchor', a 'positive', and a 'negative' for its loss calculation~\cite{triplet}.
The anchor $a$ is augmented as detailed in section~\ref{sec: Data augmentation} to obtain a positive sample $p$, while the negative sample $n$ is simply a different image in the current batch.
Using a suitable distance metric, the loss aims to minimize the distance between the anchor and positive image, while maximizing the distance between the anchor and negative image.

Triplet loss enables our encoder to embed semantic features of our marked image. Encouraging similarity between the anchor and the positive sample helps distill the essential features of the marked image and the embedded watermark into an $ID$, with the negative sample acting as a failsafe to prevent network collapse during training.

\vspace{-1.5em}
\begin{equation}
  L_{enc} = max(0, mse(M_{a}, M_{p}) - mse(M_{a}, M_{n}) + m).
  \label{eq:triplet_eq}
\end{equation}
\vspace{-1.5em}

The embedder and extractor are trained as a pair to create identical pairs of $C$ and $M$, and $W$ and $W'$. For this purpose, we employ pixel-wise mean squared error (MSE) as a distance metric between two images to train the model. 
This is illustrated in equations~\ref{eq:emb_loss} and~\ref{eq:ext_loss}:

\vspace{-1.0em}
\begin{equation}
  L_{emb} = mse(C, M),
  \label{eq:emb_loss}
\end{equation}
\vspace{-1.5em}

\vspace{-1.5em}
\begin{equation}
  L_{ext} = mse(W, W').
  \label{eq:ext_loss}
\end{equation}
\vspace{-1.5em}

The encoder's objective is to create an invariant domain (ID) for a given marked image. It is trained using triplet loss~\cite{triplet} as depicted in equation~\ref{eq:triplet_eq}, where the margin $m$ is an adjustable hyperparameter controlling the difficulty of convergence.

Lastly, the decoder and extractor are jointly trained using MSE between the input watermark triplets $W_{a}$, $W_{p}$, $W_{n}$ and the extracted watermark triplets $W_{a\_e}$, $W_{p\_e}$, $W_{n\_e}$, as illustrated in equation~\ref{eq:ext_final_loss}:

\vspace{-1.5em}
\begin{equation}
  L_{ext} = mse(W_{a}, W_{a\_e}) + mse(W_{p}, W_{p\_e}) + mse(W_{n}, W_{n\_e}).
\label{eq:ext_final_loss}
\end{equation}
\vspace{-1.5em}

\subsection{Marked Image Augmentation}
\label{sec: Data augmentation}   
As illustrated in Figure~\ref{fig:Architecture}, a specific augmentation is applied to $M_{t}$, enhancing the robustness of the watermarking scheme. Each marked image undergoes a compound augmentation process, where a variety of noises are probabilistically selected and introduced to the image. This augmentation strategy is crucial for ensuring the adaptability and resilience of the watermarking scheme against potential distortions.

As detailed in table~\ref{table:noises}, the training noises chosen for this augmentation process include horizontal flip, Gaussian blur, solarization, brightness adjustment, contrast variation, as well as hue and saturation modulation. These particular types of noises were selected based on their proven efficacy in Self-Supervised Learning (SSL) representation learning as documented in the literature~\cite{DINO}. By integrating these noises, we aim to foster a more discriminative and robust representation learning environment, which in turn, contributes to the overall effectiveness and reliability of the proposed watermarking scheme.

Furthermore, the probabilistic selection of these noises ensures a diverse range of augmentations, which simulates a variety of real-world conditions the marked images may encounter. This diversity in augmentation is instrumental in preparing the watermarking scheme to tackle a broad spectrum of challenges, thereby enhancing its practical utility and robustness in real-world applications.


\begin{table} [!ht]
\centering
\begin{tabular}{ccc}
\hline
Noise & Training & Testing\\
\hline
Horizontal flip & \bluetick & \redcross \\
Gaussian blur & \bluetick & \redcross \\ 
Solarization & \bluetick & \redcross \\ 
Crop & \redcross & \bluetick \\ 
Cutout & \redcross & \bluetick \\ 
JPEG compression & \redcross & \bluetick \\
Brightness & \bluetick & \redcross \\ 
Contrast & \bluetick & \redcross \\ 
Hue & \bluetick & \redcross \\ 
Saturation & \bluetick & \redcross \\ 
Histogram equalization & \redcross & \bluetick \\ 
Salt and pepper & \redcross & \bluetick \\ 
Gaussian noise & \redcross & \bluetick \\ 
\hline
\end{tabular}
\vspace{0.5em}
\caption{Training and testing noises.}
\label{table:noises}
\end{table}
\vspace{-1.5em}

The selected testing noises include cropping, cutout, JPEG compression, histogram equalization, salt and pepper, and Gaussian noise. These were chosen based on their recognition as the most common watermarking attacks in previous literature~\cite{facebook_DINO, Hinton}. Experiments are conducted to assess the robustness of our model against each of these noises individually.

\section{Experiments and Analysis}
\label{sec:Experiments}
This section provides an experimental analysis through quantitative and analytical evaluations of the proposed watermarking scheme.
Sections~\ref{sec: Datasets} and~\ref{sec: Data augmentation} detail the datasets utilized and the preprocessing conducted on them, along with a discussion on the augmentation and the specifics of training and testing noises.
Section~\ref{sec: Functionality and Robustness} validates the functionality of the proposed watermarking scheme and uses heavy distortions to demonstrates the robustness.
Section~\ref{sec: Ablation Study for cross attention and ID} compares the embedding fidelity between our cross attention embedder and traditional CNN-based embedding scheme and delves into an ablation study comparing the watermarking performance with and without the inclusion of the invariant module. 
Lastly, Section~\ref{sec: Comparison Experiments} juxtaposes the performance of the proposed watermarking scheme against other state-of-the-art schemes.

\vspace{-0.5em}
\subsection{Datasets}
\label{sec: Datasets}

\subsubsection{Cover image generation} 
The ImageNet dataset comprises $14,197,122$ annotated images. We utilize a highly regarded subset of ImageNet: The ImageNet Large Scale Visual Recognition Challenge (ILSVRC) 2012-2017 image classification and localization dataset to generate our cover images. Our selected subset encompasses $34,752$ training images and $3,936$ testing images, all resized to $128\times128$. We opt for this size to maintain a relatively comparable image dimension with most state-of-the-art watermarking papers.

\subsubsection{Watermark generation} 
Our goal is to generate a watermark with a 64-bit capacity. Initially, we resize a cover image to $8\times8\times1$. Subsequently, the image is binarized to zeros and ones by thresholding each pixel using a threshold value of 128. Hence, for each training or testing batch, our watermark is derived from a reshaped, binarized random image from the same batch. We repeat this process once more for our shuffled watermark as depicted in figure~\ref{fig:Architecture}.


   

\subsection{Robustness}
\label{sec: Functionality and Robustness}
This section validates the performance of the proposed watermarking scheme in extracting watermarks from heavily distorted images, showcasing its invariance and robustness to extreme distortions. The embedding fidelity is evaluated using the peak-signal-to-noise ratio (PSNR), which measures the similarity between the marked and cover images. The bit recovery rate (BRR) is used to evaluate the performance of our watermark extraction, calculated by examining the bit-by-bit similarity between two images.

\begin{figure}[!ht]
    \centering
    \vspace{-1.0em}
    \includegraphics[width=0.75\linewidth]{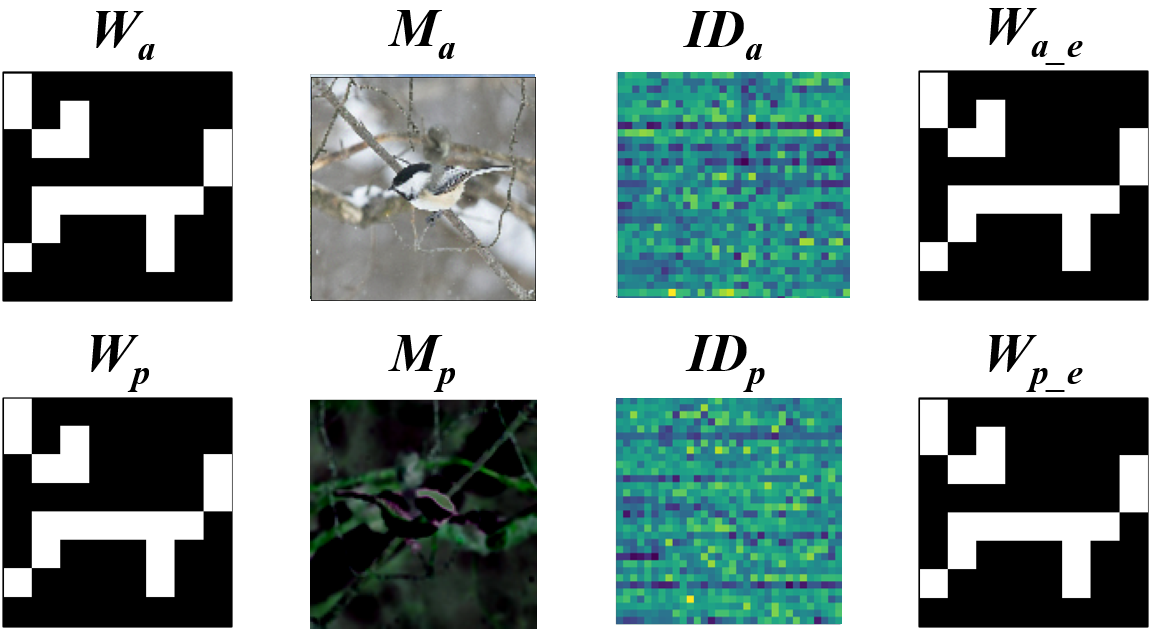}
    \vspace{-0.5em}
    \caption{Example validations of invariance in our scheme amid distortions.}
    \vspace{-0.5em}
    \label{fig:functionality_results}
\end{figure}


Figure~\ref{fig:functionality_results} visualizes some results. $W_{a}$ and $W_{p}$ represent anchor and positive watermarks, with their respective marked images $M_{a}$ and $M_{p}$, and the invariant domains obtained as $ID_{a}$, $ID_{p}$. The last column displays the extracted watermarks $W_{a\_e}$ and $W_{p\_e}$ from each invariant domain. Our results also showcase the scheme's performance on marked images under extreme augmentation, achieving a 100\% BRR in the first case and 85.93\% in the second case under heavy augmentation. 
Testing on $3,936$ ImageNet testing set images yields a BRR of 99.99\% for unaugmented and 77.39\% for augmented images using all the noises in our training noise set.

\begin{figure*}[!ht]
    \centering
    \vspace{-1.0em}
    \includegraphics[width=1.0\linewidth]{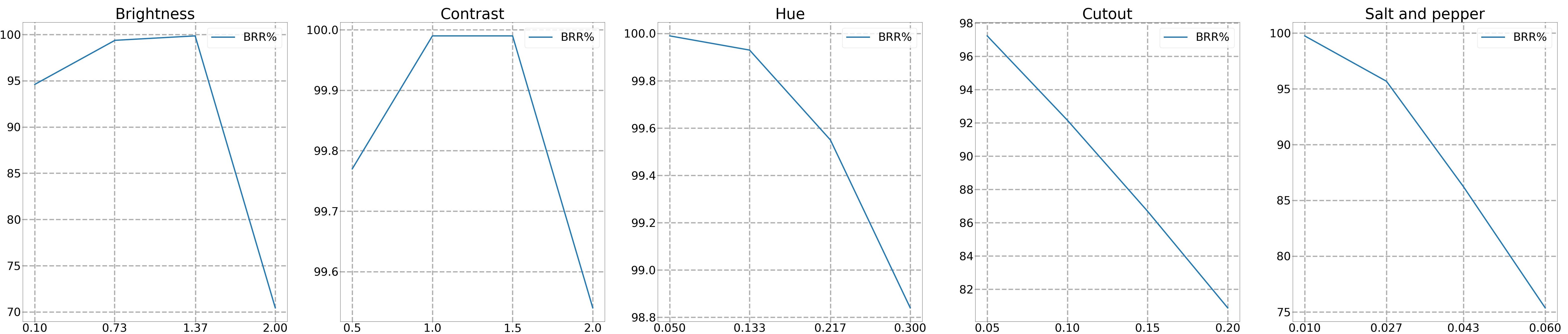}
    \vspace{-0.5em}
    \caption{Noise tolerance test of watermarking extraction with increasing noise levels.}
    \vspace{-1.0em}
    \label{fig:noise_tolerance}
\end{figure*}

In addition, we conduct experiments to evaluate the tolerance of our proposed scheme against escalating levels of noise. This test aims to gauge the robustness of our watermark extraction mechanism. 
As illustrated in Figure~\ref{fig:noise_tolerance}, a discernible trend is observed where the BRR diminishes with increasing noise intensity for both training and testing noises, aligning with the anticipated behavior. However, a notable aspect is the substantial degree of tolerance exhibited by our scheme even amidst escalating noise levels. 

The retained performance amid extreme noise highlights our proposed scheme's robustness, crucial for real-world scenarios where watermarked images may face degradation and attacks. The accurate watermark recovery under such conditions attests to our scheme's practical viability. Furthermore, the gradual performance decline with increasing noise levels, rather than abrupt failure, adds a reliability margin, essential in applications requiring high data integrity and resilience against adversarial manipulations.


\begin{figure}[!ht]
    \centering
    \vspace{-1.5em}
    \includegraphics[width=0.75\linewidth]{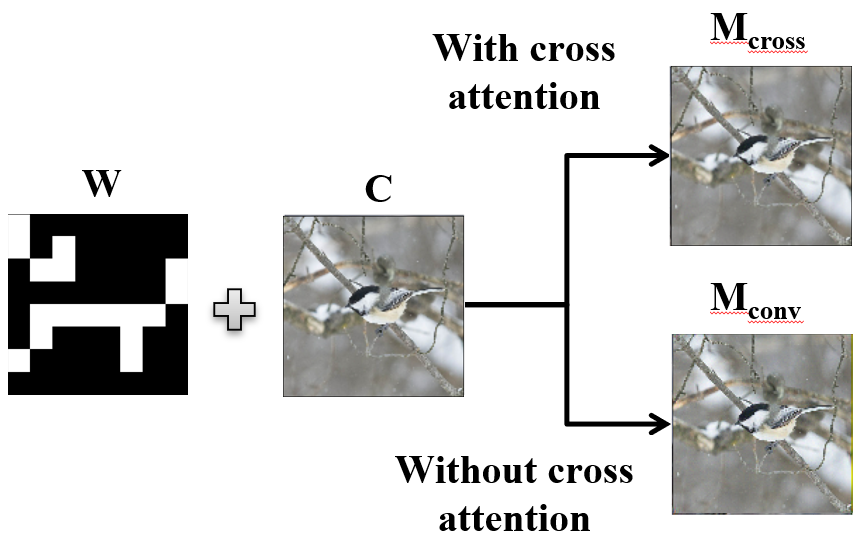}
    \vspace{-0.5em}
    \caption{Comparison of embedding fidelity with and without cross attention. 
    }
    \vspace{-1.0em}
    \label{fig:ablation_cross_results}
\end{figure}

\begin{figure}[!ht]
    \centering
    \vspace{-0.75em}
    \includegraphics[width=0.6\linewidth]{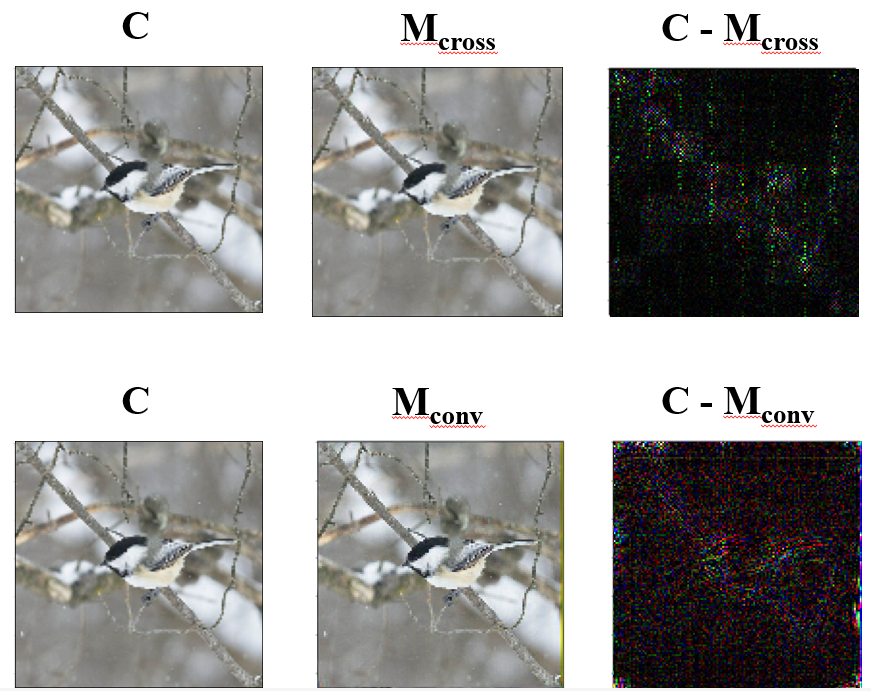}
    \vspace{-1.0em}
    \caption{Study comparing embedding fidelity with and without cross attention.
    }
    \label{fig:watermark_location}
    \vspace{-1.0em}
\end{figure}

\subsection{Ablation Study}
\label{sec: Ablation Study for cross attention and ID}

\subsubsection{Cross Attention}
We first highlight the effectiveness of our embedder that utilizes cross attention by comparing it with a CNN-based embedder. Figure~\ref{fig:ablation_cross_results} depicts an example of a watermark $W$ being embedded into a cover image $C$ both with and without our cross attention embedder. Although the difference between $M_{cross}$ and $M_{conv}$ is not immediately visually apparent, the PSNR between $C$ and $M_{cross}$ is 44.64dB, while the PSNR between $C$ and $M_{conv}$ is 32.10dB.

Over our Imagenet testing set of $3,936$ unaugmented images and generated watermarks, the average PSNR without cross attention is only 30.76dB with a BRR of 99.49\%, whereas the PSNR with cross attention is 40.94dB with a BRR of 99.57\%. This experiment demonstrates that the proposed cross attention-based embedder outperforms the CNN-based embedders.

Moreover, figure~\ref{fig:watermark_location} illustrates the pixels affected by the embedding process for each case by showcasing the difference between the cover image and their respective marked image. As the brighter pixels denote change, they represent pixels that have been modified to embed the watermark. This figure suggests that the watermark is more localized in the case of cross attention embedding compared to CNN-based embedding.

\subsubsection{Invariant Domain}
We also demonstrate the effectiveness of our proposed scheme by comparing the watermark extraction performance with and without the invariant domain, particularly in the presence of noise. Initially, we evaluate the performance of only our pretrained embedder and extractor using an augmented image and watermark pair. Subsequently, we repeat the experiment with our invariant domain and compare the difference in extraction BRR.

\begin{figure}[!ht]
    \centering
    \vspace{-1.5em}
    \includegraphics[width=0.6\linewidth]{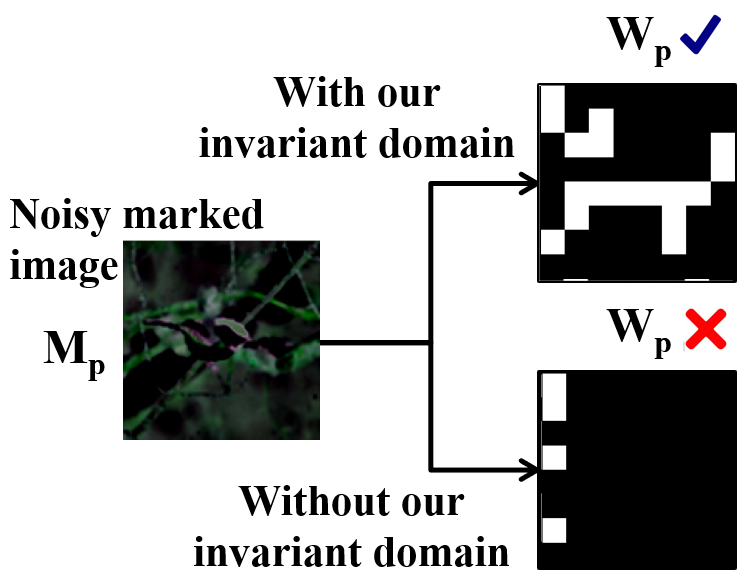}
    \caption{Ablation study comparing watermark extraction with and without our proposed invariant domain.}
    \label{fig:ablation_results}
    \vspace{-1.0em}
\end{figure}

Figure~\ref{fig:ablation_results} illustrates one example where we augment the marked image $M_{t}$ with randomly selected training noises to obtain $M_{p}$, and then attempt to recover the watermark from it. 
The lower output $W_{p}$ is an example without our invariant domain, clearly failing to extract the watermark. Averaging over our testing set of $3,936$ images, the BRR is only 45.74\%. 
Conversely, performing the same extraction from $M_{p}$ with our invariant domain results in a BRR of 100\% over our testing set (see the upper $W_{p}$ as one example). This experiment underscores that the proposed invariant domain is crucial for accurate watermark extraction from heavily noisy images.

\subsection{Comparative Analysis}
\label{sec: Comparison Experiments}
In this section, we present a comparative analysis of our watermark extraction performance against other cutting-edge approaches outlined in~\cite{Zhong}~\cite{facebook_DINO}~\cite{Distortion_agnostic}. These methodologies were specifically chosen due to the diversity in their augmentation strategies, allowing for a broader comparison across varying types of noises. As articulated in section~\ref{sec: Data augmentation}, our training noises were meticulously selected to foster a robust invariant domain, while our testing noises were chosen based on their prevalence in common watermarking attacks. We employ our proposed scheme to conduct validation experiments against both sets of noises, comparing it with four other methods. Our model is trained using an ADAM optimizer with a decaying learning rate of 0.0001.






\begin{table}[!ht]
\centering
\begin{tabular}{cccccc}
\hline
Method & BRR\% & BRR\% & BRR\% & BRR\% & BRR\%\\
Noise & (HE) & (Cutout) & (S\&P) & (JPEG) & (Blur)\\
Range & N/A & 20\% & 5\% & 10 & 2.0\\
\hline
\cite{Zhong} & 99.57 & 100 & 99.03 & 91.84 & 100\\
\textbf{Ours} & 99.67& 99.56& 97.14& 47.92& 99.53\\
\hline
\end{tabular}
\vspace{0.5em}
\caption{Comparison~\cite{Zhong}: Comparison of BRR}
\vspace{-1.5em}
\label{table:Comparison_zhong}
\end{table}

\begin{table}[!ht]
\centering
\begin{tabular}{cccccc}
\hline
Method & BRR\% & BRR\% & BRR\% & BRR\% & BRR\%\\
Noise & (Bright.) & (Contr.) & (Hue) & (JPEG) & (Blur)\\
Range & 2.0 & 2.0 & 0.25 & 50 & 2.0\\
\hline
\cite{facebook_DINO} & 91.3 & 91.6 & 97.5 & 79.2 & 99.5\\
\textbf{Ours} & 94.63& 99.79& 99.15& 48.62& 99.53\\
\hline
\end{tabular}
\vspace{0.5em}
\caption{Comparison~\cite{facebook_DINO}: Comparison of BRR}
\vspace{-1.5em}
\label{table:Comparison_meta}
\end{table}

\begin{table}[!ht]
\centering
\begin{tabular}{ccccc}
\hline
Method & BRR\% & BRR\% & BRR\% & BRR\%\\
Noise & (GN) & (Hue) & (JPEG) & (Blur)\\
Range & 0.06 & 0.2 & 50 & 1.0\\
\hline
\cite{Distortion_agnostic} & 95.6 & 94.0 & 81.7 & 92.8\\
\textbf{Ours} & 75.83& 99.15& 48.62& 96.83\\
\hline
\end{tabular}
\vspace{0.5em}
\caption{Comparison~\cite{Distortion_agnostic}: Comparison of BRR}
\vspace{-2.5em}
\label{table:Comparison_distortion}
\end{table}





Tables~\ref{table:Comparison_zhong},~\ref{table:Comparison_meta}, and~\ref{table:Comparison_distortion} elucidate the performance comparison of our proposed scheme against the selected state-of-the-art methods. The results showcase that our scheme holds its ground against other methods in terms of extraction BRR, with particular robustness against brightness, contrast, hue, histogram equalization, cutout and Salt \& Pepper (S\&P) adversarial manipulations.

We conjecture that the observed performance dip in JPEG compression scenarios may be attributed to the integration of ViT into our architecture, a phenomenon also noted in~\cite{facebook_DINO}. This conjecture invites further investigation into the interplay between ViT and JPEG compression, potentially opening avenues for further optimizations to enhance the robustness of our scheme under such adversarial conditions.



   
   
   


\section{Conclusion}
\label{sec:Conclusion}
Deep learning-based watermarking has emerged as a promising enhancement over traditional methodologies. By autonomously adapting to various watermarking requirements and distortions, these models offer a potential pathway to more robust and efficient solutions. 

The global perspective provided by ViTs has hinted at more coherent and effective image watermarking, especially where a comprehensive understanding of the image content is critical for intelligent and inconspicuous watermark embedding. Yet, the exploration of ViTs' attention capability for image watermarking remains untouched. Similarly, the concurrent training of an invariant domain alongside image watermarking has not been delved into, as prevailing methods often treat SSL models as ready-made.

In this work, we introduced a robust image watermarking method based on cross-attention and invariant domain learning, opening two notable novel avenues. Initially, we devised a watermark embedding technique using a multi-head cross attention mechanism, enabling an information interchange between the cover image and watermark to identify semantically suitable embedding locations. We then proposed learning an invariant domain that encapsulates both semantic and noise-invariant information concerning the watermark, achieved through a self-supervised watermarking framework that learns watermarking and the invariant domain concurrently.
This exploration in training methodology lays a solid foundation for enhancing robustness in deep learning-based image watermarking.
Empirical evaluations show our method either matches or outperforms the robustness across varied noise scenarios compared to state-of-the-art techniques, suggesting a promising direction for advancing deep learning-based image watermarking and encouraging further exploration into leveraging attention mechanisms and SSL paradigms.




\bibliographystyle{IEEEtran}
\bibliography{bibliography}

\end{document}